Angular Momentum of Twisted Radiation from an Electron in Spiral Motion

M. Katoh[1,2]*, M. Fujimoto[1,2], H. Kawaguchi[3], K. Tsuchiya[4], K. Ohmi[4], T. Kaneyasu[5], Y. Taira[6], M. Hosaka[7], A. Mochihashi[7], Y. Takashima[7]

[1]Institute for Molecular Science, National Institutes of Natural Sciences, Okazaki 444-8585, Japan
[2]Sokendai (the Graduated University for Advanced Studies), Okazaki 444-8585, Japan
[3]Muroran Institute of Technology, Muroran, 050-0071 Japan
[4]High Energy Accelerator Research Organization (KEK), Tsukuba, 305-0801 Japan
[5]Saga Light Source, Tosu, 841-0005 Japan
[6]National Institute of Advanced Industrial Science and Technology (AIST), Tsukuba, 305-8568 Japan
[7]Nagoya University, Nagoya, 464-0814 Japan

ABSTRACT
We theoretically demonstrate for the first time that a single free electron in circular/spiral motion emits twisted photons carrying well defined orbital angular momentum along the axis of the electron circulation, in adding to spin angular momentum. We show that, when the electron velocity is relativistic, the radiation field contains harmonic components and the photons of $l$-th harmonic carry $l\hbar$ total angular momentum for each. This work indicates that twisted photons are naturally emitted by free electrons and more ubiquitous in laboratories and in nature than ever been thought.

Twisted photons that carry an orbital angular momentum (OAM) in addition to a spin angular momentum (SAM) have been intensively studied for applications in information, nano- and imaging technologies[1]. Their interactions with atoms[2], molecules[3], materials[4,5] and plasmas[6] are also being explored. A photon with OAM was originally discussed with respect to a specific mode of electromagnetic wave called the Laguerre-Gaussian mode[7], and the discussion was later extended to more general cases[8]. It was shown that, when the radiation field has a phase term represented by $e^{in\phi}$, each photon carries an OAM of $n\hbar$, where $\phi$ is the azimuthal angle around the propagation axis, $n$ is an integer, and $\hbar$ is Planck constant.

These previous works addressed mathematical models for the twisted electromagnetic waves without considering the radiation processes themselves. Under the requirement that their models satisfy the Helmholtz equations, the authors discussed the existence of AM. In the laboratories, researchers have developed various technologies to produce electromagnetic waves of the Laguerre-Gaussian mode. Currently, twisted photons in the visible and infrared wavelength ranges can be readily produced using conventional laser sources and holographic filters[1]. On the other hand, there is much less effort to seek physical processes that produce twisted photons in nature. Some authors have reviewed the possible roles of twisted photons in astrophysics[9,10], but they mainly discussed how to detect the twisted photons instead of how to create them. Other authors have addressed the modifications of radiation fields by a gravitational field around a rotating black hole[11] or inhomogeneous interstellar media[12] away from the radiation processes.

Here, we show theoretically for the first time that electromagnetic wave radiated by a single free electron in circular motion has a helical phase structure and carries OAM. This is in contrast with the works in the past[7,8], which demonstrated that an electromagnetic wave carrying OAM can exist and propagate. This work demonstrates that such an electromagnetic wave can be radiated by a charged particle. This work will bring the vortex photon science to a new stage where the vortex photons are discussed in the context of real physical processes, how they are created, how they propagate and what role they play in laboratory systems and in natural systems.

The radiation from an electron in circular motion was first addressed by O.

Heaviside in 1904[13]. It is the basis of the radiation processes by electrons in a magnetic field or an intense circular polarized light field, which are known as cyclotron/synchrotron radiation or Thomson/Compton scattering. Because they play important roles in astrophysics, plasma physics and accelerator physics, the radiation from an electron in circular/spiral motion has been addressed in many textbooks[14,15] and literature[16,17]. However, surprisingly, no study has discussed its phase structure or angular momentum for more than one hundred years. In this paper, we treat this radiation process paying a special attention to the phase structure of the radiation field. We derive its vector potential in a spherical coordinate. Then, we demonstrate its helical phase structure and its AM.

Generally, the real part of the time-averaged AM density $\vec{j}$ of an electromagnetic field can be represented as a cross product of the linear momentum density $\vec{p}$ and the position vector $\vec{r}$ as follows[18]:

$$\langle \vec{j} \rangle = \vec{r} \times \langle \vec{p} \rangle = \frac{1}{4\pi c} \vec{r} \times (\vec{E} \times \vec{H}^*) = \frac{r}{4\pi c} \left( \vec{E}(\vec{n} \cdot \vec{H}^*) - \vec{H}^*(\vec{n} \cdot \vec{E}) \right) \quad (1)$$

Here, $c$ is the light velocity, $\vec{E}$ and $\vec{H}$ are the electric and magnetic fields, the asterisk on the shoulder represents their complex conjugate, and $\vec{n}$ is the unit vector that directs the observer, which is defined as $\vec{n} = \vec{r}/r$. The brackets $\langle \ \rangle$ indicate time-average. In the radiation zone, the electric and magnetic fields are approximately perpendicular to $\vec{n}$ [5]. In this case, the inner products $\vec{n} \cdot \vec{H}^*$ and $\vec{n} \cdot \vec{E}$, and therefore AM, are approximately zero. It should be noted that an electromagnetic field with AM should have a non-zero component along the direction of propagation.

By integrating Eq. (1) on a spherical surface that surrounds the source, the AM that the radiation field carries away can be expressed as[15]:

$$\frac{d\vec{J}}{dt} = \int c \langle \vec{j} \rangle r^2 d\Omega = \int r^3 d\Omega \frac{1}{4\pi} \left( \vec{E}(\vec{n} \cdot \vec{H}^*) - \vec{H}^*(\vec{n} \cdot \vec{E}) \right) \quad (2)$$

Here, $\Omega$ is the solid angle, and the integration is taken over a spherical surface at distance $r$ from the origin. The energy that the radiation carries away can be expressed as[15]:

$$\frac{dU}{dt} = \int r^2 d\Omega \frac{c}{4\pi} \left( \vec{E} \times \vec{H}^* \right) \quad (3)$$

We can see in Eq. (3) that for the energy flow, it is sufficient to consider terms in the integrant up to the order of $1/r^2$ because the higher-order terms disappear at a sufficiently large distance ($r \to \infty$). However, for AM, we must consider the terms up to the order of $1/r^3$.

We treat the electron motion and radiation field in a coordinate system shown in Fig. 1. An electron travels on a circular orbit around the origin. Here, we do not assume the cause of this circular motion. The electron motion can be specified by only two parameters: its velocity $\beta$ and its angular frequency $\omega$. The radius $r_e$ of the motion can be expressed with these parameters as $r_e = c\beta/\omega$. Although the direction of electron motion is another parameter, for simplicity, here, we select it as counterclockwise around the z-axis as shown in Fig. 1. We treat this problem in the spherical coordinate in contrast with previous works[7,8], where the cylindrical coordinate was selected. Generally, radiation propagates in a free space as a spherical wave; therefore, it seems best described in the spherical coordinate.

The electromagnetic field radiated by a moving electron can be represented by a retarded potential. When the electron motion is periodic, it can be decomposed into Fourier series as follows:[14]

$$\vec{A}(\vec{r},t) = e \left. \frac{\vec{\beta}}{|\vec{r}-\vec{r}_e| - (\vec{r}-\vec{r}_e)\cdot\vec{\beta}} \right|_{t_e} = \sum_{l=1}^{\infty} \vec{\hat{A}}_l(\vec{r}) e^{-il\omega t} \qquad (4)$$

Here, $\vec{r}$ is the position vector of the observing point, $\vec{r}_e$ and $\vec{\beta}$ are the electron position and velocity, respectively, as previously described, which should be expressed with the electron time $t_e = t - |\vec{r}-\vec{r}_e|/c$. It should be noted that when the electron velocity is much smaller than the light velocity, only the fundamental component exists. When the electron velocity is comparable to the light velocity, e.g., it is relativistic, the harmonic components appear.

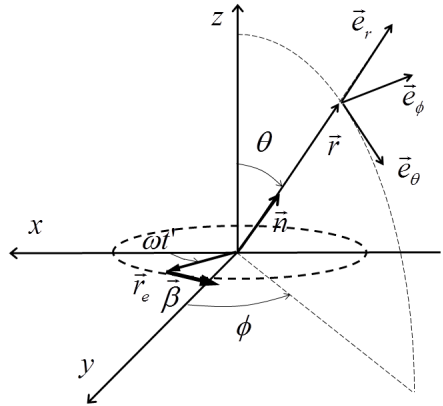

Figure 1. Coordinate System. The azimuthal angle $\phi$ is taken from the y-axis.

The Fourier components of the vector potential can be expressed as follows:[14]

$$\widehat{\vec{A}}_l(\vec{r}) = \frac{e}{c}\frac{1}{2\pi/\omega}\frac{e^{ikr}}{r}\oint e^{il\omega(t_e - \frac{\vec{n}\cdot\vec{r}_e(t_e)}{c})} d\vec{r}_e(t_e) + o\left(\frac{1}{r^2}\right) \quad (5)$$

Here, we separate the term with the first order of $1/r$ from the other terms for convenience in later discussion. The position vector of the electron can be represented in the spherical coordinate as follows:

$$\vec{r}_e = r_e\{\sin\theta\sin(\omega t_e - \phi)\vec{e}_r + \cos\theta\sin(\omega t_e - \phi)\vec{e}_\theta - \cos(\omega t_e - \phi)\vec{e}_\phi\} \quad (6)$$

By inserting Eq. (6) into Eq. (5), we can obtain the following expression:

$$\vec{A}_l(\vec{r},t) = \begin{pmatrix} A_{lr} \\ A_{l\theta} \\ A_{l\phi} \end{pmatrix} = \frac{e}{c}\frac{e^{i\frac{l\omega}{c}r}}{r}e^{il\phi}r_e\omega \begin{pmatrix} \sin\theta\frac{1}{2\pi}\oint e^{i(l\varphi - l\beta\sin\theta\sin\varphi)}\cos\varphi d\varphi \\ \cos\theta\frac{1}{2\pi}\oint e^{i(l\varphi - l\beta\sin\theta\sin\varphi)}\cos\varphi d\varphi \\ \frac{1}{2\pi}\oint e^{i(l\varphi - l\beta\sin\theta\sin\varphi)}\sin\varphi d\varphi \end{pmatrix} + \widehat{\vec{A}}_l^{(2)} + \cdots$$

$$= e\frac{e^{i(kr - l\omega t + l\phi)}}{r}\begin{pmatrix} J_l(l\beta\sin\theta) \\ \cot\theta J_l(l\beta\sin\theta) \\ i\beta J_l'(l\beta\sin\theta) \end{pmatrix} + o\left(\frac{1}{r^2}\right) \quad (7)$$

Here, we introduced a new parameter $\varphi = \omega t_e - \phi$ and express the integration in Eq. (5) by Bessel functions of the 1st order. We also used the relation $k \equiv l\omega/c$.

The electric and magnetic fields can be expressed by the vector potential as follows:

$$\begin{pmatrix} H_{lr} \\ H_{l\theta} \\ H_{l\phi} \end{pmatrix} = \begin{pmatrix} 0 \\ -\frac{1}{r}\frac{\partial}{\partial r}(rA_{l\phi}^{(1)}) \\ \frac{1}{r}\frac{\partial}{\partial r}(rA_{l\theta}^{(1)}) \end{pmatrix} + \begin{pmatrix} \frac{1}{r\sin\theta}\frac{\partial}{\partial\theta}(\sin\theta A_{l\phi}^{(1)}) - \frac{1}{r\sin\theta}\frac{\partial}{\partial\phi}A_{l\theta}^{(1)} \\ + o\left(\frac{1}{r^2}\right) \\ + o\left(\frac{1}{r^2}\right) \end{pmatrix} + o\left(\frac{1}{r^3}\right) \quad (8)$$

$$\begin{pmatrix} E_{lr} \\ E_{l\theta} \\ E_{l\phi} \end{pmatrix} = \begin{pmatrix} 0 \\ \frac{1}{r}\frac{\partial}{\partial r}\left(rA_{l\theta}^{(1)}\right) \\ \frac{1}{r}\frac{\partial}{\partial r}\left(rA_{l\phi}^{(1)}\right) \end{pmatrix} + \begin{pmatrix} -\frac{1}{r\sin\theta}\frac{\partial}{\partial\theta}\left(\sin\theta A_{l\theta}^{(1)}\right) - \frac{1}{r\sin\theta}\frac{\partial}{\partial\phi}A_{l\phi}^{(1)} \\ o\left(\frac{1}{r^2}\right) \\ o\left(\frac{1}{r^2}\right) \end{pmatrix} + o\left(\frac{1}{r^3}\right) \qquad (9)$$

Here, we present the results by separating the terms of the order of $1/r$. In Eqs. (8) and (9), the first bracket on the right hand side is the first order of $1/r$, and the second bracket is the second order. Later, we represent those terms as $E^{(i)}$ and $H^{(i)}$, and the index on the shoulder represents the order of $1/r$. Some components are not explicitly shown because they are not necessary in the following discussion. It should be noted that in the order of $1/r^2$, we can find non-zero electric and magnetic field components in the propagation direction, which would contribute to AM in Eq. (2). In addition, all necessary field components in the following discussion can be expressed by the derivatives of the lowest-order terms of the vector potential in Eq. (7). Before calculating AM, we will show the helical phase structure of the radiation field. The electric and magnetic fields in the lowest order of $1/r$ can be derived from Eqs. (8) and (9) as follows:

$$\begin{pmatrix} E_{lr}^{(1)} \\ E_{l\theta}^{(1)} \\ E_{l\phi}^{(1)} \end{pmatrix} = \begin{pmatrix} 0 \\ \frac{1}{r}\frac{\partial}{\partial r}\left(rA_{l\theta}^{(1)}\right) \\ \frac{1}{r}\frac{\partial}{\partial r}\left(rA_{l\phi}^{(1)}\right) \end{pmatrix} = \begin{pmatrix} 0 \\ ikA_{l\theta}^{(1)} \\ ikA_{l\phi}^{(1)} \end{pmatrix} = \frac{e}{c}l\omega\begin{pmatrix} 0 \\ i\cot\theta J_l(l\beta\sin\theta) \\ -\beta J_l'(l\beta\sin\theta) \end{pmatrix}\frac{e^{i(kr-l\omega t+l\phi)}}{r} \qquad (10)$$

To clearly show the phase structure, we express these terms in the Cartesian coordinate as follows:

$$\begin{pmatrix} E_{lx}^{(1)} \\ E_{ly}^{(1)} \\ E_{lz}^{(1)} \end{pmatrix} = \begin{pmatrix} -\sin\theta\sin\phi & -\cos\theta\sin\phi & -\cos\phi \\ \sin\theta\cos\phi & \cos\theta\cos\phi & -\sin\phi \\ \cos\theta & -\sin\theta & 0 \end{pmatrix}\begin{pmatrix} 0 \\ E_{l\theta}^{(1)} \\ E_{l\phi}^{(1)} \end{pmatrix} \qquad (11)$$

The final result can be expressed as a sum of circular polarized components of positive and negative helicities, which are proportional to the rotation vectors defined as:

$$\vec{E} = E^{(1)}_{lx}\vec{e}_x + E^{(1)}_{ly}\vec{e}_y + E^{(1)}_{lz}\vec{e}_z = \frac{\left(E^{(1)}_{lx} - iE^{(1)}_{ly}\right)}{\sqrt{2}}\frac{\vec{e}_x + i\vec{e}_y}{\sqrt{2}} + \frac{\left(E^{(1)}_{lx} + iE^{(1)}_{ly}\right)}{\sqrt{2}}\frac{\vec{e}_x - i\vec{e}_y}{\sqrt{2}} + E^{(1)}_{lz}\vec{e}_z$$

$$\equiv E^{(1)}_{l+}\vec{e}_+ + E^{(1)}_{l-}\vec{e}_- + E^{(1)}_{lz}\vec{e}_z \tag{12}$$

$$= \frac{e}{c}l\omega\frac{e^{i(kr-l\omega t)}}{r}\begin{pmatrix} \left\{\frac{\cos^2\theta}{\sin\theta}J_l(l\beta\sin\theta) + \beta J_l'(l\beta\sin\theta)\right\}e^{i(l-1)\phi}\vec{e}_+ \\ +\left\{-\frac{\cos^2\theta}{\sin\theta}J_l(l\beta\sin\theta) + \beta J_l'(l\beta\sin\theta)\right\}e^{i(l+1)\phi}\vec{e}_- \\ -i\cot\theta J_l(l\beta\sin\theta)e^{il\phi}\vec{e}_z \end{pmatrix}$$

As shown in Eq. (12), the electric field can be decomposed into two circular polarized components with respect to the z-axis in the positive and negative helicities with the phase terms $e^{i(l-1)\phi}$ and $e^{i(l+1)\phi}$, respectively. We will discuss this phenomenon later.

By inserting Eqs. (8) and (9) to Eqs. (2) and (3), the AM density can be expressed as:

$$\langle\vec{j}_l\rangle = \frac{r}{4\pi c}\left(0 \quad E^{(1)}_{l\theta}H^{(2)*}_{lr} - H^{(1)*}_{l\theta}E^{(2)}_{lr} \quad E^{(1)}_{l\phi}H^{(2)*}_{lr} - H^{(1)*}_{l\phi}E^{(2)}_{lr}\right)_{r,\theta,\phi} + o\left(\frac{1}{r^3}\right) \tag{13}$$

From the symmetry of the system, the projected AM to the z-axis is expected to be well defined[15]:

$$\langle j_{lz}\rangle = -\langle j_{l\theta}\rangle\sin\theta = \frac{r}{4\pi c}\sin\theta\left(E^{(2)}_{lr}H^{(1)*}_{l\theta} - E^{(1)}_{l\theta}H^{(2)*}_{lr} + o\left(\frac{1}{r^4}\right)\right) \tag{14}$$

The total angular momentum carried away by the radiation can be obtained by integrating Eq. (14) on the sphere surrounding the source as follows:

$$\left\langle\frac{dJ_{lz}}{dt}\right\rangle = \int cr^2\langle j_{lz}\rangle d\Omega \tag{15}$$

By inserting Eqs. (8) and (9) to Eq. (14) and using Eq. (15), we obtain:

$$\left\langle\frac{dJ_{lz}}{dt}\right\rangle = \int d\Omega\frac{cr^2}{4\pi c}\left[kl\left(A^{(1)}_{l\theta}A^{(1)*}_{l\theta} + A^{(1)}_{l\phi}A^{(1)*}_{l\phi}\right) - ik\left\{A^{(1)}_{l\theta}\frac{\partial}{\partial\theta}\left(\sin\theta A^{(1)*}_{l\phi}\right) + A^{(1)*}_{l\phi}\frac{\partial}{\partial\theta}\left(\sin\theta A^{(1)}_{l\theta}\right)\right\}\right]$$

$$= \frac{r^2}{4\pi}kl\int d\Omega\left(A^{(1)}_{l\theta}A^{(1)*}_{l\theta} + A^{(1)}_{l\phi}A^{(1)*}_{l\phi}\right) \tag{16}$$

Here, the terms in the brace in Eq. (16) vanish in the integration regarding to $\theta$.

On the other hand, the energy that the electromagnetic wave carries away can be obtained from the Poynting vector, whose radial component is

expressed as follows:

$$\langle S_{lr} \rangle = \left\langle \frac{c}{4\pi} \vec{E}_l \times \vec{H}_l^* \right\rangle_r = \frac{c}{4\pi}\left(E_{l\theta}^{(1)}H_{l\phi}^{(1)*} - E_{l\phi}^{(1)}H_{l\theta}^{(1)*}\right) = \frac{c}{4\pi}k^2\left(A_{l\theta}^{(1)}A_{l\theta}^{(1)*} + A_{l\phi}^{(1)}A_{l\phi}^{(1)*}\right) \quad (17)$$

Then, the carried energy is;

$$\left\langle \frac{dU_l}{dt} \right\rangle = \frac{ck^2}{4\pi}\int r^2 d\Omega\left(A_{l\theta}^{(1)}A_{l\theta}^{(1)*} + A_{l\phi}^{(1)}A_{l\phi}^{(1)*}\right) \quad (18)$$

Using Eqs. (16) and (18), the ratio of AM to the energy that the radiation field carries is obtained as:

$$\frac{\left\langle \frac{dJ_{lz}}{dt} \right\rangle}{\left\langle \frac{dU_l}{dt} \right\rangle} = \frac{\frac{kl}{4\pi}\int r^2 d\Omega\left(A_{l\theta}^{(1)}A_{l\theta}^{(1)*} + A_{l\phi}^{(1)}A_{l\phi}^{(1)*}\right)}{\frac{ck^2}{4\pi}\int r^2 d\Omega\left(A_{l\theta}^{(1)}A_{l\theta}^{(1)*} + A_{l\phi}^{(1)}A_{l\phi}^{(1)*}\right)} = \frac{l}{ck} = \frac{l}{l\omega} = \frac{N_p \hbar l}{N_p \hbar l \omega} \quad (19)$$

Here, we have introduced Plank constant $\hbar$ and the number of photons $N_p$ as in the previous works[7,8]. Eqs. (19) clearly show that the radiation field carries non-zero AM. Moreover, it is consistent with a quantum mechanical picture that each photon carries $l\hbar$ AM. Here, the SAM and OAM are not clearly separated, in contrast with a previous work adopting the paraxial approximation[7], but in common with another work adopting the non-paraxial approximation[8].

We have shown that the ratio of the AM density in the z-direction to the energy density is expressed in a simple form in Eqs. (19). Previous works assumed mathematical models for the radiation field, where the polarization was treated as a free parameter. However, in our case, we began from the electron motion, which is counterclockwise as previously described. Therefore, the polarization is defined by the electron motion and not a free parameter. We see the polarization of the radiation field from Eq. (12), which is mostly circular polarized when the polar angle $\theta$ is small. The direction of the polarization is identical to the electron motion (we will call this direction positive for convenience). When $\theta$ increases, the polarization becomes elliptical. In other words, the field becomes a mixture of circularly polarized components with positive and negative helicities. When $\theta$ is $\pi/2$, it is linearly polarized, which implies that the intensities of both helicity components are identical. When $\theta$ is larger than $\pi/2$, the negative helicity components becomes dominant. It should be noted that the positive helicity component has the phase term represented by $e^{i(l-1)\phi}$ and the negative

$e^{i(l+1)\phi}$. We speculate that, if we consider the z-component of AM, in the former case, SAM is +1 and OAM is $l-1$, and in the latter case, SAM is -1, and OAM is $l+1$. In any case, the total AM is always $l$.

It may be instructive to consider the angular momentum in the non-linear inverse Compton process of circularly polarized light, which is also an example of the radiation from an electron in circular motion. In classical mechanics, intense circularly polarized light induces a circular motion of an electron. When the light is sufficiently intense, the electron motion becomes relativistic, and higher harmonics are radiated. In quantum mechanics, this process can be described as a multi-photon process. Let us consider that $n$ photons with the energy of $\hbar\omega$ are coming to the electron. Because they are circularly polarized, they bring $\hbar$ AM for each; therefore, they bring $n\hbar$ AM to the system in total. One photon with energy $n\hbar\omega$ is outgoing, which implies that the energy is conserved. However, the outgoing photon can bring at most $\pm\hbar$ AM as SAM. For the conservation of angular momentum, it should carry the remaining AM as another form of AM, which is OAM. If the outgoing photon carries $\hbar$ as SAM, then it should carry other $(n-1)\hbar$ as OAM. When it carries $-\hbar$ as SAM, then it should carry $(n+1)\hbar$ as OAM. This speculation seems consistent with the above discussion on Eq. (12).

It should be noted that the ratio of the AM density to the energy density does not depend on the explicit form of the vector potential as observed in Eq. (19). We think that this fact assures the Lorentz invariance of the angular momentum. As we remind the derivation, Eqs. (16) and (18) should be valid for any vector potential with the following form:

$$\vec{A}_l(\vec{r},t) = e\frac{e^{i(kr-l\omega t+l\phi)}}{r}\vec{a}_l(\theta) + o\left(\frac{1}{r^2}\right) \ . \tag{20}$$

In the Lorentz transformation, a spherical wave should be transformed to a spherical wave. Indeed, by using Eq. (7), it is straightforward to obtain the vector potentials for the radiation from an electron in spiral motion that drifts along the z-axis at a relativistic speed:

$$A_{lr} = e\frac{e^{i(kr-l\omega t+l\phi)}}{r}\frac{1}{1-\beta_z\cos\theta}J_l\left(l\frac{K}{\gamma}\frac{\sin\theta}{1-\beta_z\cos\theta}\right)+o\left(\frac{1}{r^2}\right)$$

$$A_{l\theta} = e\frac{e^{i(kr-l\omega t+l\phi)}}{r}\frac{\cos\theta-\beta_z}{1-\beta_z\cos\theta}\frac{1}{\sin\theta}J_l\left(l\frac{K}{\gamma}\frac{\sin\theta}{1-\beta_z\cos\theta}\right)+o\left(\frac{1}{r^2}\right) \quad (21)$$

$$A_{l\phi} = e\frac{e^{i(kr-l\omega t+l\phi)}}{r}\frac{iK/\gamma}{1-\beta_z\cos\theta}J_l\left(l\frac{K}{\gamma}\frac{\sin\theta}{1-\beta_z\cos\theta}\right)+o\left(\frac{1}{r^2}\right)$$

Here, $\beta_z$ is the electron velocity along the z-axis, $\gamma$ is Lorentz factor, and $K/\gamma$ is the pitch angle of the electron motion to the z-axis. The angular frequency $\omega$ is related to that in the electron frame $\omega^*$ as follows:

$$\omega = \frac{\sqrt{1+K^2}}{\gamma(1-\beta_z\cos\theta)}\omega^* \quad (22)$$

Eq. (21) shows that the vector potentials in the laboratory frame have the identical form to those in Eq. (20). Therefore, Eqs. (18) and (19) are valid for this case, and the AM in the z-direction is assured to be invariant to the Lorentz transformation. Using Eq. (21), one can derive the electromagnetic field for a helical undulator[18,19], inverse Compton scattering of circular polarized light[20] or cyclotron/synchrotron radiation[16,17] with the phase terms which represent their vortex nature.

In this paper, we have theoretically shown that the photons radiated by an electron in a spiral motion are twisted and carry OAM. As far as we know, this is the first theoretical demonstration that radiation field by a single charged particle carries well defined OAM. This process is one of the most fundamental radiation processes in laboratories and in nature. Therefore, we conclude that photons with OAM are much more ubiquitous than previously thought. They may play unexplored, important roles with respect to their OAM. In addition, this process will be the basis of laboratory vortex photon sources that cover the entire wavelength range from radio waves to gamma rays, and they will open completely new research opportunities.

Acknowledgments;
MK contributed to the entire derivation of the analytic formulas with helps by HK, KT and KO on the mathematical approaches. The analytic formulas were checked by MF, TK, MH, AM, Y Tak and Y Tai. The manuscript was prepared by MK based on the discussions with all others and was checked by all others. This work was partially supported by JSPS KAKENHI Grant Number 26286081.